\documentclass[runningheads]{llncs}
\usepackage{cite}
\usepackage{amsmath,amssymb,amsfonts}
\usepackage{algorithmic,subfigure}
\usepackage{graphicx}
\usepackage{textcomp}
\usepackage[misc]{ifsym}
\usepackage{xcolor}
\begin{document}
\title{A Derivative-free Method for Quantum Perceptron Training in Multi-layered Neural Networks}
\vspace{-1.2cm}
\titlerunning{Derivative-free Quantum Perceptron Training}
\author{Tariq M. Khan {\Letter}%\inst{1}\orcidID{0000-0002-7477-1591} \and
\ \ \ \ \ \ \ \ \ \ \ \ \ \ \ \ \
Antonio Robles-Kelly%\inst{1}\orcidID{0000-0002-2465-5971}
}
\institute{Deakin University, Faculty of Sci., Eng. and the Built Env.\\
Waurn Ponds, VIC 3216, Australia\\
\email{tariq.khan@deakin.edu.au, antonio.robles-kelly@deakin.edu.au}
%\email{Paper ID: 342}
}
\tocauthor 
\toctitle
\maketitle              % typeset the header of the contribution
\vspace{-0.8cm}
\begin{abstract}
In this paper, we present a gradient-free approach for training multi-layered neural networks based upon quantum perceptrons. Here, we depart from the classical perceptron and the elemental operations on quantum bits, {\it i.e.} qubits, so as to formulate the problem in terms of quantum perceptrons. We then make use of measurable operators to define the states of the network in a manner consistent with a Markov process. This yields a Dirac–Von Neumann formulation consistent with quantum  mechanics. Moreover, the formulation presented here has the advantage of having a computational efficiency devoid of the number of layers in the network. This, paired with the natural efficiency of quantum computing, can imply a significant improvement in efficiency, particularly for deep networks. Finally, but not least, the developments here are quite general in nature since the approach presented here can also be used for quantum-inspired neural networks implemented on conventional computers.
\vspace{-0.2cm}
\end{abstract}
\vspace{-1.2cm}
\section{Introduction}
\vspace{-0.4cm}

%Quantum computation is an emerging technology that uses the quantum-mechanical phenomena to perform computations.
Quantum computing algorithms often exhibit significant increases in efficiency, in some cases exponentially, compared to their classical counterparts. This is particularly relevant to machine learning, which has had a growing importance in recent years. This is since machine learning methods tend to be computationally intensive. Thus, recently, there have been numerous research studies aiming to investigate the promise of quantum computers for machine learning \cite{Lloyd2014,Jeswal2019,Biamonte2017,Chen2017}.

Moreover, it has been suggested that quantum computers may be an ideal platform for the implementation of artificial neural networks \cite{Biamonte2017}. In recent years ongoing attempts have been made to implement artificial neural networks (ANN) in quantum computers.
Grover et al. \cite{Grover96} attempted to emulate quantum computation on  classical computers to perform search quadratically faster than its classic equivalent in an unordered dataset. The computational power of quantum computing in terms of efficiency and effectiveness for ANNs as compared to that of classical computers has also been explored in \cite{Jeswal2019}. Nonetheless, the incorporation of quantum computation into ANNs is still an open and challenging research direction \cite{Chen2017}.

This is further compelled by the fact that deep learning is an algorithmic class within the wider category of machine learning algorithms with their own practical and architectural properties. Deep nets are used primarily to classify patterns on a specific data set and/or to produce new data that imitates these patterns. At heart, there are three main components in neural network algorithms. Firstly, the model, comprised by a parametric functional hypothesis class, typically set up in a network of layered composition of simpler parametric functions. %Multi-layer networks are called deep and are the subclass of deep-learning models.
Secondly, a cost function, which determines how well the prediction based upon the input data fits a specific hypothesis. Thirdly, the optimizer. This is an algorithmic technique used to minimise the loss function based upon the parameters in the network. This is often done by backward error propagation, also known as the backpropagation algorithm. This is at the core of ANN training.

Moreover, the cost (error) function of ANNs is often purely a function of the network's output. %Sometimes a function of the output of certain network subsets.
The backpropagation algorithm is the most common method in both, quantum and classical computing to train ANNs\cite{Wang2012}. This is used to train the network using the cost function gradient (in relation to the network parameters), beginning with the output layer going layer-by-layer towards the input one. In 1986, the backpropagation algorithm was proposed by Rumelhart and Mcllelland to solve the non-linear continuous function weight adjustment problems in the area of the neural multi-layer feedforward network as a back error method \cite{Rumelhart1986}.

Since the development of the back-propagation method for neural networks, a lot of research has been carried out on the choice of activation function, design of structure parameters and characterise the loss function. A lot of research has also been carried out to improve the efficiency of the back-propagation methods in neural networks. Sun et al. \cite{Sun2014} have developed an improved prediction model of back-propagation neural network and quantitatively researched related parameters. Xiao {\it et al.} \cite{Xiao2009} present a short-term load forecast method for Neural Network Prediction.
%The accuracy of prediction of back-propagation neural network was further improved \cite{Xiao2009}.

Here we note that, although the back-propagation algorithm is the most widely used one in artificial neural networks in both classical as well as quantum domains, it does exhibit the following drawbacks:

\begin{itemize}
  \item It can fall into local extremum points if there are multiple points on the loss space with zero gradient \cite{Zhang2009}. %Also, if the error need to rise as a part of a more general fall, the algorithm will get stuck and in such a scenario the error will not decrease further.
  \item Its speed of convergence is dependent on two aspects. Firstly, the learning rate. Secondly, the magnitude of the gradient associated to the excitation function \cite{Xu2015}. This can result in slow convergence rates. This is further complicated by the fact that the magnitude of the learning rate is not a straightforward parameter to set. If the learning rate is too large, the back propagation solver will often suffer from ``over shooting''. If the learning rate is too small, the network may fail to converge at all.
  \item Its computational complexity is dependent on the network structure, increasing with the number of hidden layers and the number of neurons per layer.
\end{itemize}

%In conclusion, backpropagation isn’t perfect as it require a lot of memory, is difficult to parallelize and can't apply directly to non-differentiable functions. A viable alternate to backpropagation is a gradient free approach which treats the problem like a black-box and calculate its fitness. It doesn't give any potentially incorrect assumptions and many non-differentiable functions that are not possible to implement in propagation can be implemented in ES. As, in ES, no backpropagation required therefore, these are embarrassingly parallel and require much less memory than backpropagation.

In this paper, we turn out attention to the evaluation of the quantum perceptron as a means to tackle the drawbacks above in the training of neural networks. To do this, we depart from the concept of a single quantum bit, {\it i.e.} a qubit, and examine the strategies for evaluating the equivalent in quantum computing of the perceptron in machine learning. We then focus on measurable operators to model the evolution of an artificial neural network as the action of a unitary transformation on the network's present firing states. This has the advantage that, for training an artificial neural network, the forward pass is given by a measurement of the transformed state whereas the training can be effected using a gradient-free strategy whose complexity is devoid of the number of layers in the network. This, together with the natural efficiency of quantum neural nets, is a promising trait that can greatly speed up both training and testing of neural networks in quantum computing. This unitary transformation would, of course, have to include information on the probabilities of transition between each basic state.

\vspace{-0.3cm}
\section{Background}
\vspace{-0.3cm}

In this section, we briefly introduce the concepts of quantum computing that are necessary for the remainder of the paper. For a detailed introduction on quantum computing, we would like to remit the interested reader to \cite{Biamonte2017}.

\vspace{-0.3cm}
\subsection{Quantum Bits}
\vspace{-0.2cm}

As mentioned earlier, we depart from the concept of qubit. A quantum bit or qubit can be represented by a linear combination of two base states using the ``Bra-ket'' notation, {\it i.e.} the pairing of a linear function and complex vector in a Hilbert space, as follows  $\left| 0 \right\rangle  = \left[ 1,0 \right]^T$ and $\left| 1 \right\rangle  = \left[ 0, 1\right]^ T$, where, as usual, $[\cdot]^T$ is a column vector given by the transpose of a row vector. These are used to define the qubit as $ \left| \psi  \right\rangle  = a\left| 0 \right\rangle  + b\left| 1 \right\rangle  = \left[ \psi _0,\psi _1\right]^T$.

In the expression above, ${\psi _0}$ and ${\psi _1}$ are complex numbers, usually called probability amplitudes in the literature. This treatment leads, in a straightforward manner to the natural extension to a multi-qubit expression. To represent a system with multiple qubits, a tensor product, which yields a matrix representation can be employed without any loss of generality. To illustrate this, consider, for instance, the two-qubit case where the quantum bits are given by $\left| \psi  \right\rangle  = \left[ \psi _0,\psi _1\right]^T$ and $\left| \phi  \right\rangle  = \left[\phi _0,\phi _1\right]^T$. Using this notation, their tensor product is given by $\left| \psi  \right\rangle  \otimes \left| \phi  \right\rangle  = \left[
\psi _0\phi _0,
\psi _0\phi _1,
\psi _1\phi _0,
\psi _1\phi _1 \right]^T$.
From this product, it becomes evident that the resultant output is a four-dimensional vector. Moreover, this can be extended to any pair of vector in an $m$ and $n$ dimensions, which would yield yet another $(m\times n)$-dimensional vector.

\vspace{-0.2cm}
\subsection{Classical Perceptron}
\vspace{-0.2cm}

Recall that a perceptron is a binary classification algorithm for supervised learning, which is the simplest type of neural network. Let the $i^{th}$ instance of the dataset be the input vector $\mathbf{x}_i$ % = [x_1,x_2,...,x_N]^T$
to the perceptron. The output of the perceptron is based upon the vector $\mathbf{y}$ whose $j^{th}$ entry is governed by $\langle\mathbf{w}_j,\mathbf{x}_i\rangle+b_j$, where $b_j$ is a constant, {\it i.e.} the bias, $\mathbf{w}_j$ is a vector of weights and $\langle\cdot,\cdot\rangle$ is the dot product as usual. The output of the perceptron is then given by a function $ y_i=f(\mathbf{y})$ where each vector of weights $\mathbf{w}_j$  corresponds to a neuron. Thus, the output  for the input instance $\mathbf{x}_i$ is given by
\begin{equation}
y_i = f\bigg( {\sum\limits_{j = 1}^N \langle\mathbf{w}_j,\mathbf{x}_i\rangle+b_j } \bigg)
\end{equation}
where $f\left(  \cdot  \right)$ is known as the activation function, $\mathbf{W}$ is a matrix of weights whose $j^{th}$ row corresponds to $\mathbf{w}_j^T$ and $\mathbf{b}$ is a vector whose entry indexed $j$ is given by $b_j$.

\vspace{-0.3cm}
\section{Derivative-free Training of a Quantum Perceptron}
\vspace{-0.3cm}\[ \]

As mentioned earlier, the qubit is often represented using the Bra-ket notation. In this manner, following the notion that $\mathbf{x}_i$ can be represented using the notation $\left| {{x_1}} \right\rangle$, consider  the training set written in the form $\left\{ {\left( {\left| {{x_1}} \right\rangle ,\left| {{y_1}} \right\rangle } \right), \ldots ,\left( {\left| {{x_N}} \right\rangle ,\left| {{y_N}} \right\rangle } \right)} \right\}$ where ${\left| {{x_j}} \right\rangle }$ is an input and ${\left| {{y_j}} \right\rangle }$ is the corresponding label.

Recall that, in order to obtain the weight vector ${\left| {{{ w}_j}} \right\rangle }$, as suggested in \cite{8631025}, a tensor product can be used, which yields $ \mathrm{W}_i = \left| y_i \right\rangle  \otimes \left\langle x_i \right|$. After calculating the vector of weights using the expression above, these can be added to get the final weight vector given by $\mathrm{\hat W} = \sum_{i = 1}^N \mathrm{W}_i$. Note that the matrix $\hat W$ is, in general, not unitary. Thus, to preserve the quantum properties requiring a unitary matrix, $\hat w$ is decomposed into three unitary matrices using the singular value decomposition (SVD) given by $\mathbf{\hat W} = \mathbf{U}\mathbf{\Sigma}\mathbf{V}^*$.
%\begin{equation}\label{10}
% \hat F\sum {{{\hat w}_{new}}}
%\end{equation}

As noted by Liu {\it et al.} \cite{8631025}, the diagonal matrix $\\mathbf{\Sigma}$ can be substituted, without any loss of generality, with a unitary matrix with ones in diagonal and zeros elsewhere, {\it i.e.} and identity matrix. This yields, using the notation commonly employed in quantum machine learning texts where the unitary matrix $\mathbf{U}$ is denoted by $\mathrm{\hat F}$, the quantum perceptron output given by
$\mathrm{\hat Y} = \mathrm{\hat F}%\sum\nolimits_{new} {{{\hat w}_{new}}}
\lvert x_j \rangle$. Here, we follow Zak and Wiliams \cite{Zak1998}, who viewed an $n$-neuron network as a dynamic system that obeys the differential equation given by
 \begin{equation}\label{6}
{\tau _i}\frac{\partial }{{\partial t}}\mathrm{Z}_i =  - \mathrm{Z_i} + f\left( {\sum\limits_{j = 1}^N {\mathrm{W_j}\mathrm{Z_j}} } \right)
\end{equation}
where $\tau _i$ is a positive time constant, $\mathrm{Z}_i$ is the activation of the $i^{th}$ neuron and $\mathrm{W}_j$ are synaptic weights analogue to those elaborated upon previously that feed the activation function of the neuron indexed $j$ to the activation function $f(\cdot)$.

Equation \ref{6} can be used to formulate the update of the quantum perceptron weights as follows
\begin{equation}
\label{7}
\mathrm{W}^{new}  =  M\left\{ \mathrm{U}\mathrm{\hat Y} - \mathrm{W}^{old} \right\}
\end{equation}
where $\mathrm{W}^{old}$ is the current, {\it i.e.} old, weights, $\mathrm{U}$ is a unitary matrix that acts on the state vector $\mathrm{\hat Y}$, $M$ is a measurable operator that project states of $\mathrm{U}\mathrm{\hat Y}$ into some eigenstate of $M$. The use of measurable operators naturally provides a link to statistics and quantum measurements. The equation \ref{7} shows the update of the perceptron weights by using the state vector projected upon U (the orthonormal basis spanned by the SVD of the sum of outer products in $\mathrm{\hat W} = \sum_{i = 1}^N \mathrm{W}_i$).

%\begin{figure*}
 % \centering
 % \includegraphics[scale=1.3]{block.eps}
 % \caption{Block diagram of evaluation strategies based neural network.Note that $\mathrm{Q}_i=\sum\limits_{j = 1}^N \langle\mathbf{w}_j,\mathbf{x}_i\rangle$ in classical computing. }\label{block}
%\end{figure*}

\vspace{-0.3cm}
\section{Measurable Operators}
\vspace{-0.3cm}

In this section, we examine closer the role of the the measurable operator $M$. This is due to the fact that it can open-up several opportunities as related to the design of neural networks, specially as related to approaches elsewhere in the literature that employ backpropagation methods for training. Note that  $ \mathrm{U}\mathrm{\hat Y}$ can be viewed as a representation of a sequence of measurable states that define a Markov process with a transition probability matrix. Moreover, Quantum probability is a non-commutative extension of classical probability which represents random variables as self-adjoint operators that act on a complex Hilbert space whereby the underlying probability is measured by a unitary vector.

Furthermore, note that $M$ is, by definition, a self-adjoint operator. In quantum mechanics, self-adjoint operators form a Dirac–Von Neumann formulation of quantum mechanics. Self-adjoint operators represent the physical observables such as spin, momentum, angular momentum and position on a Hilbert space. Recall that, by definition, probability distributions are non-negative and normalised to unity. Since self-adjoint operators are are unitarily equivalent to real-valued multiplication operators, they can be easily generalised to theoretically unbounded operators on infinite-dimensional spaces. Here, we will also require $M$ to be hermitian. This is important since then, by definition, has an orthonormal set of eignevectors $\lvert \xi  \rangle$ with real eigenvales  ${\lambda _i}$.

Note that, if the elements of $\mathrm{U}$ are appropriately chosen, then any desired Markov chain can be simulated without a random number generator. This is possible due to the inherent randomness of quantum measurement processes. The modulus square of the probability amplitude represents the probability of a network in the $j^{th}$ transitions to the $i^{th}$ state. This can be expressed using the following expression
\begin{equation}\label{12}
\Pr \left[ \mathrm{Y}^{new}   = \lvert i \rangle \mathrm{Y}^{old}  =\lvert j \rangle \right] = \lvert\lvert u_{ij} \lvert\lvert^2
\end{equation}
where $u_{ij}$ is the entry indexed $i,j$ of $\mathrm{U}$ and we have used the notation $\mathrm{Y}^{new}$ and $\mathrm{Y}^{old}$ to denote the current and previous states of the network. It is worth noting in passing that, since
$\mathrm{U}$ is a unitary matrix, there will be appropriate constraints on the possible values of $\lvert\lvert u_{ij} \lvert\lvert^2$ so as to avoid divergent behavior. We remit the interested reader to \cite{Zak1998} for further reading on this.
\vspace{-0.3cm}
\section{Algorithm Properties}
\vspace{-0.2cm}
\subsection{Complexity}
\vspace{-0.2cm}

It is worth noting that the complexity of the computations above will be far more efficient in a quantum computer than running an equivalent simulation on a classical computer. The reason being that quantum computers have the potential to be exponentially larger than classical computers, with the capacity to represent random stochastic states naturally \cite{Zak1998}.

Also, note the SVD is exponentially  faster in quantum computers than on classical ones \cite{PhysRevA.97.012327}. Rebentrost {\it et. al.} \cite{PhysRevA.97.012327} proposed quantum-SVD for non-sparse low rank matrices with complexity $O(poly \log N)$. Gyongyosi {\it et. al.} \cite{gyongyosi2012improvement} proposed an algorithm for quantum-SVD with complexity $O(N \log N)$. This complexity is similar to the standard Fourier transform complexity of $O(N \log N)$ \cite{985949}. On a classical computer, singular value decomposition of non-sparse low-rank matrices has a complexity, in general, and without further structural assumptions, of $O(N^3)$  \cite{PhysRevA.97.012327}. Further, in quantum computing, the multiplication is much faster than the classical computing. The space complexity of Karatsuba multiplication for numbers of $n$ bits in quantum spans from $O(n^{1.427})$ to $O(n)$ while maintaining a gate complexity of $O(n^{\log_2 3})$ \cite{gidney2019asymptotically}, where $log_2$ denotes the binary logarithm. This is achieved by avoiding the need to store and compute intermediate results \cite{gidney2019asymptotically}.

In terms of the actual quantum speedup, recall this is often either exponential or strong exponential \cite{Papageorgiou_2013}. In exponential quantum speedups, a quantum computer can solve a problem exponentially faster than a classical algorithm based upon the computational cost of the best known classical algorithm. In strong exponential quantum speedups, however, the reference is the classical complexity of the method itself. The method presented here can be measured in terms of a strong exponential speedup on the SVD given by

\begin{equation}\label{COMP2}
{S_2} = \frac{{ classical\,complexity}}{{quantum\,complexity}}=\frac{O(N^3)}{O(N \log N)}=O\bigg(\frac{N^2}{\log N}\bigg)
\end{equation}

%according to criterion,
%
%\begin{equation}\label{COMP1}
%{S_1} = \frac{{cost\,of\,the\,best\,known\,classical\,algorithm\,}}{{cost\,of\,a\,quantum\,algorithm}}
%\end{equation}

Note that the main difference between exponential and strong exponential speedups resides in the fact that computational cost is ever decreasing due to efficiency increments in memory, data structures, etc. For strong exponential speedups, it can always be asserted that a quantum computer can solve a problem exponentially faster than a classical algorithm according. This is important since quantum feedback networks are exponentially faster than those on classical computers, but they can never attain a strong exponential quantum speedup \cite{kerenidis2019quantum}. This leaves as the only avenue to obtain strong exponential quantum speedup in quantum neural network as that of using evaluation strategies with strong exponential speedups. Moreover, despite quantum neural networks employing back-propagation methods can take  advantage of these faster multiplication and efficient integration requirements, the evaluation strategy presented earlier replaces multiplication with a subtraction operation in its backward step. This is a significant improvement with respect to backpropagation, specially for large, very deep networks.

To appreciate this more clearly, recall that, in quantum computing, if the network consists of only one perceptron, then the complexity of our evaluation strategy presented earlier is consistent with that of back-propagation. This only applies for the one perceptron case since, when the number of hidden layers of the network increases, then the complexity of the back-propagation will increase significantly. On the other hand, the complexity of the evaluation strategy presented earlier will not increase with respect to the number of layers in the network. Therefore, the developments presented earlier  open-up the door for the efficient evaluation and training of complex, multi-layered networks that can be prohibitively costly computationally with back-propagation approaches. As quantum computing also provides space-complexity advantages over classical methods, increasing the number of hidden layers does not impose large memory constraints. For instance, a quantum associative memory has an exponential gain in storage capacity as compared to classical associative memories \cite{Trugenberger2001ProbabilisticQM}.
\vspace{-0.3cm}
\begin{figure}[!t]
\centering
\subfigure {\includegraphics[scale=0.5]{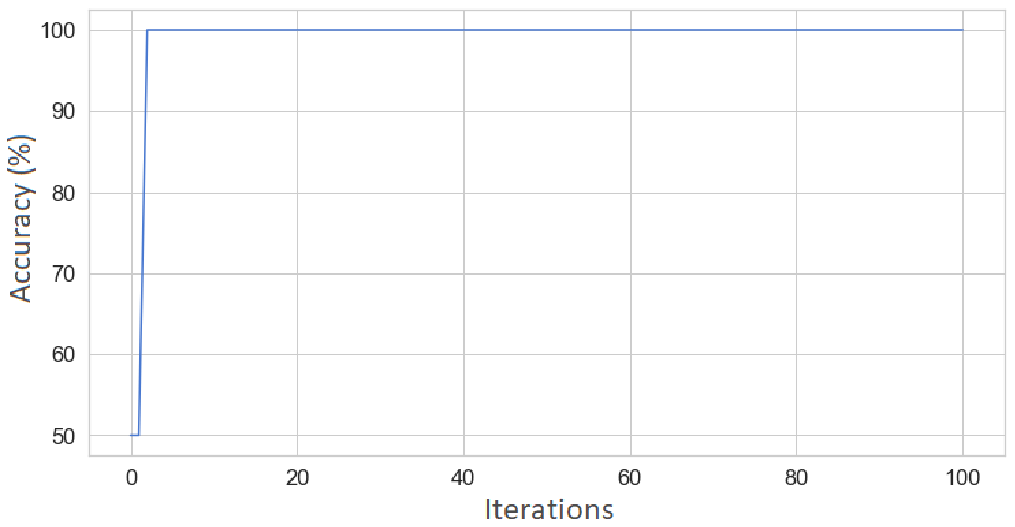}}
\subfigure {\includegraphics[scale=0.5]{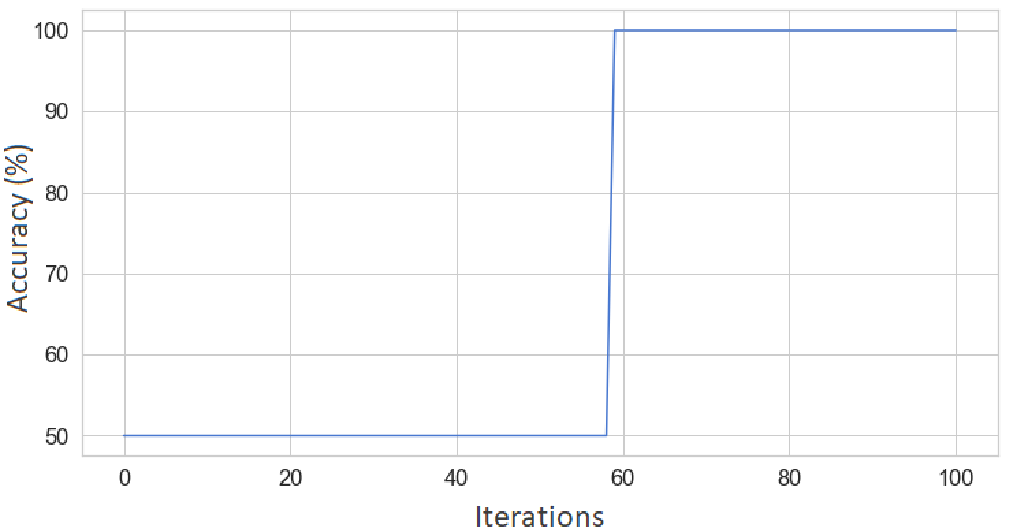}}
\subfigure {\includegraphics[scale=0.5]{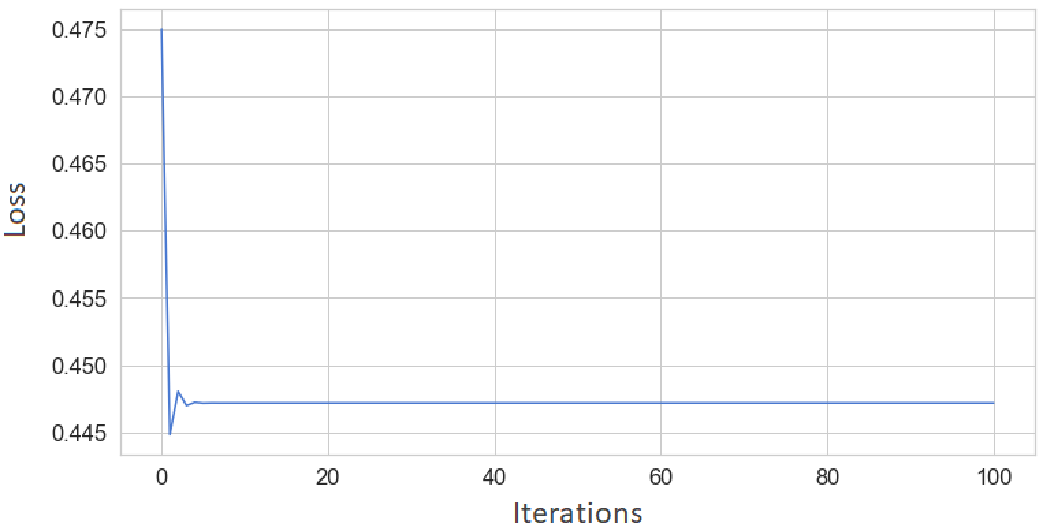}}
\subfigure {\includegraphics[scale=0.5]{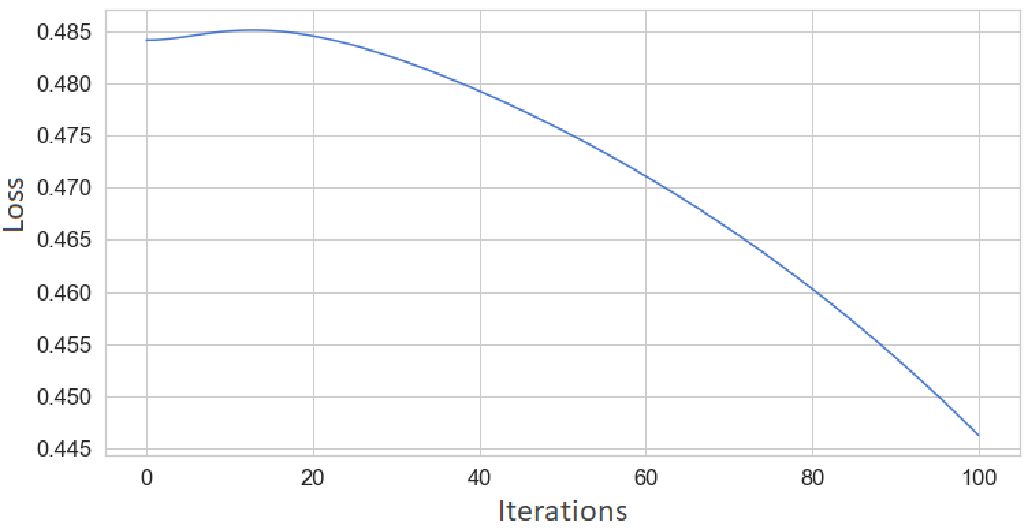}}
\centering
\vspace{-0.4cm}
 \caption{Simulation results for a perceptron trained using our approach (left-hand column) and backpropagation (right-hand column). The top row show the error for the XOr gate whereas the bottom row shows the loss. In all the plots, the independent axis corresponds to the training index.}
\label{PerformanceAnalysis}
\vspace{-0.5cm}
\end{figure}

\vspace{-0.2cm}
\subsection{Simulation Results}
\vspace{-0.2cm}

%We now turn our attention to the simulation of an XOr gate.
We have followed standard practice for the proof of concept by implementing the XOr. This is since the XOr is a universal non-linear gate which is sufficient for all logic operations on a quantum computer that can be used to construct arbitrary unitary transformations. This is a classical problem in artificial neural network research. The problem is that of, given two binary inputs, predict the output of an exclusive or gate.  An XOr gate returns a false value (a zero) if the inputs are equal and true (unity) if the inputs are not. In this case, we have used this simulation in order to illustrate the convergence rate and loss value of a quantum perceptron network with a hidden layer with two neurons and an L-$1$ loss function trained using both, our approach and backpropagation.

In order to provide a plain field for training both networks, we have employed a sigmoid as the measurable operator in Equation \ref{7}.
In our implementation in order to optimize connection weights we chose the average error as an objective function.
For measuring accuracy, $0.5$ is used as a cut-off value between zero and one. Both networks are trained for 100 iterations, whereby, at each of these, the inputs are generated randomly. At each iteration, we verify the output with another, randomly generated testing instance pair of inputs. In the top row of Figure \ref{PerformanceAnalysis} we show the error obtained using the instance input pair for both networks as a function of iteration number. The left-hand panel in Figure \ref{PerformanceAnalysis} shows that backpropagation achieves 100\% accuracy, {\it i.e.} delivers the correct XOr output for the corresponding inputs,  after 58 iterations. Our method achieves the same accuracy after 2 iterations. Moreover, in the bottom row of Figure  \ref{PerformanceAnalysis} we show the loss function for both approaches. Note that, by training using our approach, the loss converged to a local minimum after 3 epochs. On the other hand,  backpropagation didn't converged to a local minimum even after 100 epochs, as shown in Figure  \ref{PerformanceAnalysis}.

\vspace{-0.3cm}
\section{Discussion and Conclusion}
\vspace{-0.2cm}

Note that it is not unusual that millions of iterations of back-propagation may be required to train an ANN with stochastic gradient descent. For every iteration, the forward pass calculates the output of the network and, in the backward pass,  calculates the gradient with respect to the weights of the network. The network weights are then updated by an amount proportional to the gradient. This computational demands makes training ANNs one of the key drivers of increasing demand for high performance computing. This makes particularly compelling the investigation of methods for forward and backward steps that  can be speeded up by quantum computations. Moreover, derivative-free strategies in quantum computing can remove some of these steps altogether. One of the advantages of using strategies like the one presented here is that they require a forward pass but does not involve a back-propagation step. This makes these methods 2 to 3 times faster than those based upon back-propagation in a traditional computer. Moreover, an additional advantage is that, by using these evaluation strategies, a large number of non-differentiable excitation functions can be explored.

This also applies to the objective functions of quantum networks, which can potentially be discontinuous or even non-compact in neural networks trained using back-propagation methods. The strategies such as that presented here can also be very effective when solutions are known to be within an elliptical domain, {\it i.e.} around a fixed point. Moreover, gradient descent, despite effective, can be  inefficient, particularly if the  choice of starting point is poor, whereby the training can easily converges to a local minimum which may be far from an ideal solution.

Quantum-inspired neural networks (QiNNs) and Quantum computing-based neural networks have been shown to be more effective and efficient as compared to conventional ANNs \cite{Sagheer2019}. In addition, QiNN models are not limited to solely those that can only be implemented on quantum computers, rather there has been renewed interest in methods that can take advantage of QiNN traits while being implemented on conventional computers \cite{Sagheer2019}.

\vspace{-0.3cm}
 %Generated by IEEEtran.bst, version: 1.14 (2015/08/26)

\end{document}